\documentclass[final,3p,times,twocolumn]{elsarticle}
\usepackage{lineno,hyperref}
\modulolinenumbers[5]

\usepackage{amssymb}

\journal{Physics Letters B}

\begin{document}

\begin{frontmatter}

\title{Pole position of $\Lambda(1405)$ measured in $d(K^-,n)\pi\Sigma$ reactions}

%
%
%
\author{J-PARC E31 Collaboration}
\author[titech,riken1]{S.~Aikawa}
\author[rcnp]{S.~Ajimura}
\author[osaka]{T. Akaishi}
\author[riken1]{H.~Asano}
\author[victoria]{G.~Beer}
\author[smi]{C.~Berucci\fnref{fn1}}
\fntext[fn1]{current affiliation: Leonardo SpA, Rome, Italy}
\author[ifinhh]{M.~Bragadireanu}
\author[smi]{P. Buehler}
\author[torino,utorino]{L.~Busso}
\author[smi]{M.~Cargnelli}
\author[snu]{S.~Choi}
\author[frascati]{C.~Curceanu}
\author[kek1]{S.~Enomoto}
\author[titech]{H.~Fujioka}
\author[tokyo]{Y.~Fujiwara}
\author[osakaec,riken2]{T.~Fukuda}
\author[frascati]{C.~Guaraldo}
\author[asrc]{T.~Hashimoto}
\author[tokyo]{R.~S.~Hayano}
\author[riken3]{T.~Hiraiwa}
\author[kek2]{M.~Iio}
\author[frascati]{M.~Iliescu}
\author[rcnp]{K.~Inoue\corref{corrauthor}}
\ead{kentaro@rcnp.osaka-u.ac.jp}
\author[kyoto]{Y.~Ishiguro}
\author[kek3]{S.~Ishimoto}
\author[tokyo]{T.~Ishikawa}
\author[riken1]{K.~Itahashi}
\author[kek4]{M.~Iwai}
\author[riken1,titech]{M.~Iwasaki}
\author[tokyo]{K.~Kanno}
\author[kyoto]{K.~Kato}
\author[riken1]{Y.~Kato}
\author[rcnp]{S.~Kawasaki\corref{corrauthor}}
\ead{shinngo@rcnp.osaka-u.ac.jp}
\author[tum]{P.~Kienle\fnref{fn2}}
\fntext[fn2]{deceased}
\author[kek3]{Y.~Komatsu}
\author[titech]{H.~Kou}
\author[riken1]{Y.~Ma}
\author[smi]{J.~Marton}
\author[tokyo2]{Y.~Matsuda}
\author[osakaec]{Y.~Mizoi}
\author[torino]{O.~Morra}
\author[riken1]{R.~Murayama}
\author[kyoto]{T.~Nagae}
\author[rcnp,kek3]{H.~Noumi\corref{corrauthor}}
\cortext[corrauthor]{corresponding authors}
\ead{noumi@rcnp.osaka-u.ac.jp}
\author[elph]{H.~Ohnishi}
\author[riken1]{S.~Okada\fnref{fn3}}
\fntext[fn3]{current affiliation: Chubu Univ., Kasugai, Japan}
\author[rcnp,knu]{Z.~Omar}
\author[riken1]{H.~Outa}
\author[centrofermi,frascati2]{K.~Piscicchia}
\author[elph]{Y.~Sada}
\author[osaka]{A.~Sakaguchi}
\author[riken1]{F.~Sakuma}
\author[kek1]{M.~Sato}
\author[frascati]{A.~Scordo}
\author[kek3]{M.~Sekimoto}
\author[frascati]{H.~Shi}
\author[rcnp]{K.~Shirotori}
\author[frascati,ifinhh]{D.~Sirghi}
\author[frascati,ifinhh]{F.~Sirghi}
\author[smi]{K.~Suzuki}
\author[kek3]{S.~Suzuki}
\author[tokyo]{T.~Suzuki}
\author[asrc]{K.~Tanida}
\author[lund]{H.~Tatsuno}
\author[elph]{A.~O.~Tokiyasu}
\author[titech]{M.~Tokuda}
\author[rcnp]{D.~Tomono}
\author[kek3]{A.~Toyoda}
\author[elph]{K.~Tsukada\fnref{fn4}}
\fntext[fn4]{current affiliation: ICR, Kyoto Univ., Uji, Japan}
\author[frascati,tum2]{O.~Vazquez-Doce}
\author[smi]{E.~Widmann}
\author[riken1]{T.~Yamaga}
\author[riken1,tokyo]{T.~Yamazaki}
\author[kirams]{H.~Yim}
\author[riken1]{Q.~Zhang}
\author[smi]{J.~Zmeskal}
\address[titech]{Department of Physics, Tokyo Institute of Technology, Tokyo, 152-0551, Japan}
\address[rcnp]{Research Center for Nuclear Physics (RCNP), Osaka University, Ibaraki, 567-0047, Japan}
\address[osaka]{Department of Physics, Osaka University, Toyonaka, 560-0043, Japan}
\address[riken1]{RIKEN Cluster for Pioneering Research (CPR), RIKEN, Wako, 351-0198, Japan}
\address[victoria]{Department of Physics and Astronomy, University of Victoria, Victoria BC V8W 3P6, Canada}
\address[smi]{Stefan-Meyer-Institut f\"{u}r subatomare Physik, A-1030 Vienna, Austria}
\address[ifinhh]{National Institute of Physics and Nuclear Engineering - IFINHH, Romania}
\address[torino]{INFN Sezione di Torino, Torino, Italy}
\address[utorino]{Dipartimento di Fisica Generale, Universita'di Torino, Torino, Italy}
\address[snu]{Department of Physics, Seoul National University, Seoul, 151-742, South Korea}
\address[frascati]{Laboratori Nazionali di Frascati dell'INFN, I-00044 Frascati, Italy}
\address[kek1]{Accelerator Laboratory, High Energy Accelerator Research Organization (KEK), Tsukuba, 305-0801, Japan}
\address[tokyo]{Department of Physics, The University of Tokyo, Tokyo, 113-0033, Japan}
\address[osakaec]{Laboratory of Physics, Osaka Electro-Communication University, Neyagawa, 572-8530, Japan}
\address[riken2]{RIKEN Nishina Center for Accelerator-Based Science, RIKEN, Wako, 351-0198, Japan}
\address[asrc]{ASRC, Japan Atomic Energy Agency (JAEA), Ibaraki 319-1195, Japan}
\address[riken3]{RIKEN SPring-8 Center, RIKEN, Hyogo, 679-5148, Japan}
\address[kek2]{Cryogenics Science Center, High Energy Accelerator Research Organization (KEK), Tsukuba, 305-0801, Japan}
\address[kyoto]{Department of Physics, Kyoto University, Kyoto, 606-8502, Japan}
\address[kek3]{Institute of Particle and Nuclear Studies, High Energy Accelerator Research Organization (KEK), Tsukuba, 305-0801, Japan}
\address[kek4]{Mechanical Engineering Center, High Energy Accelerator Research Organization (KEK), Tsukuba, 305-0801, Japan}
\address[tum]{Technische Universit\"{a}t M\"{u}nchen, D-85748, Garching, Germany}
\address[tokyo2]{Graduate School of Arts and Sciences, The Univeristy of Tokyo, Tokyo, 153-8902, Japan}
\address[elph]{Research Center for Electron Photon Science (ELPH), Tohoku University, Sendai, 982-0826, Japan}
\address[knu]{Department of Physics, Al-Farabi Kazakh National University, Almaty, 050040, Kazakhstan}
\address[centrofermi]{Museo Storico della Fisica e Centro Studi e Ricerche ``Enrico Fermi'', Piazza del Viminale 1, 00184 Rome, Italy}
\address[frascati2]{INFN, Laboratori Nazionali di Frascati, Via Enrico Fermi 40, 00044 Frascati, Italy}
\address[lund]{Department of Chemical Physics, Lund University, Lund, 221 00, Sweden}
\address[tum2]{Excellence Cluster University, Technische Universit\"{a}t M\"{u}nchen, D-85748, Garching, Germany}
\address[kirams]{Korea Unstitute of Radiological and Medical Sciences (KIRAMS), Seoul, 139-706, South Korea}

\begin{abstract}
We measured a set of $\pi^\pm\Sigma^\mp$, $\pi^0\Sigma^0$, and $\pi^-\Sigma^0$ invariant mass spectra below and above the $\bar{K}N$ mass threshold in $K^-$-induced reactions on deuteron.  We deduced the $S$-wave $\bar{K}N\rightarrow\pi\Sigma$ and $\bar{K}N\rightarrow\bar{K}N$ scattering amplitudes in the isospin 0 channel in the framework of a $\bar{K}N$ and $\pi\Sigma$ coupled channel. We find that a resonance pole corresponding to $\Lambda(1405)$ is located at 
1417.7$^{+6.0}_{-7.4}$(fitting errors)$^{+1.1}_{-1.0}$(systematic errors) + 
$[-26.1^{+6.0}_{-7.9}$(fitting errors)$^{+1.7}_{-2.0}$(systematic errors)]$i$ MeV/$c^2$, 
closer to the $\bar{K}N$ mass threshold than the value determined by the Particle Data Group. 
\end{abstract}


\end{frontmatter}


\section{Introduction}

$\Lambda(1405)$ is a well-known hyperon resonance with strangeness $-1$, spin-parity 1/2$^-$, and isospin 0 ($I$ = 0). It is classified as the first orbital excited state in the constituent quark model. 
However, the properties of $\Lambda(1405)$ are not easily explained, such as the fact that it has the lightest mass among the negative parity baryons even though it contains a heavier strange quark, and the large mass difference it exhibits compared to that for the so-called spin--orbit partner state of $\Lambda(1520)$. 
It has been argued that $\Lambda(1405)$ is a bound state of an anti-kaon ($\bar{K}$) and a nucleon ($N$) since it is located just below the $\bar{K}N$ mass threshold, 
Dalitz and Tuan first predicted a possible quasi-bound state of $\bar{K}N$ with $I$ = 0 in 1959, based on low-energy $K^-$-proton scattering experiments \cite{dalitz59,dalitz60}. The first observation of a hyperon resonance sitting just below the $\bar{K}N$ mass threshold in $\pi^-\Sigma^+/\pi^+\Sigma^-$ invariant mass spectra was reported in 1961 \cite{lambda61}. Since then, several sets of experimental data on $\Lambda(1405)$ have been reported \cite{lambda60s,tomas,braun,hemingway,ahn,prakhov,niiyama,zychor,agakishiev,moriya13,moriya14,lu,scheluchin}.  Dalitz and Deloff deduced a resonance energy and width of 1406.5 $\pm$ 4.0 MeV and 50 $\pm$ 2 MeV by analyzing the measured $\pi^-\Sigma^+$ mass spectrum \cite{hemingway} based on $\bar{K}N$ scattering theory \cite{dalitz91}. The latest edition of the Review of Particle Physics \cite{pdg} gives average values of 1405.1$^{+1.3}_{-0.9}$ MeV and 50.5 $\pm$ 2.0 MeV, including two later works \cite{esmile,hassanvand} which demonstrate that a so-called phenomenological approach giving the $\Lambda(1405)$ mass at $\sim$1405 MeV \cite{ay1,ay2} is favorable for fitting the $(\pi\Sigma)^0$ invariant mass spectra from $K^-$ stopped on $^4$He \cite{riley75} and proton--proton collisions (HADES)  \cite{agakishiev}. A recent review on $\Lambda(1405)$ is available in Ref.~\cite{ppnp2021}.

Over the last two decades, there have been intensive discussions about the so-called chiral unitary approach, which is a coupled-channel meson--baryon scattering theory employing chiral Lagrangians. Several calculations indicate that there are two resonance poles between the $\pi\Sigma$ and $\bar{K}N$ mass thresholds \cite{oller01,jido,hyodo06,hyodo07,hyodo08}, where the higher pole, coupled to $\bar{K}N$, is located at around 1420 MeV or greater. The chiral unitary approach is in contradiction with the phenomenological approach. There is a discussion of differences in different theoretical treatments of chiral unitary approaches and the phenomenological approach \cite{morimatsu}.

The experimental situation is also controversial. 
Recent measurements of $(\pi\Sigma)^0$ mass spectra have been reported in photo-induced reactions on  protons \cite{niiyama,moriya13,moriya14,lu,scheluchin} and proton--proton collisions \cite{zychor,agakishiev}. The CLAS collaboration reported precise $\pi^-\Sigma^+$, $\pi^+\Sigma^-$, and $\pi^0\Sigma^0$ spectra for a wide range of incident photon energies \cite{moriya13,moriya14}. Theoretical analyses have been made on these data and reproduced the spectral shapes fairly well, even though they involved many parameters \cite{roca} and/or reaction diagrams \cite{nakamura}. The HADES collaboration reported invariant mass spectra of $\pi^-\Sigma^+$, $\pi^+\Sigma^-$, and their sum \cite{agakishiev}. Their spectral shapes were different from those for photo-production. In particular, they observed peaks even below 1400 MeV.  Theoretical analyses of these spectra have also been made \cite{siebenson,hassanvand}. However, the locations of $\Lambda(1405)$ determined by a chiral unitary model \cite{siebenson} and a phenomenological model \cite{hassanvand} are not compatible with each other. Therefore, experimental data to directly determine the $\bar{K}N$ scattering amplitude coupled to $\Lambda(1405)$ are required.

\section{Experiment}

\begin{figure}
\centering
\includegraphics[width=7.8cm]{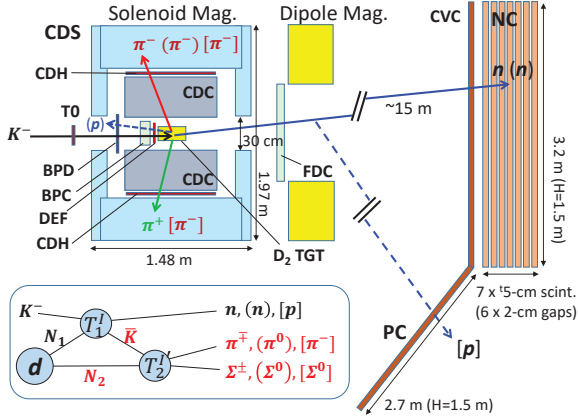}%
\caption{\label{fig:diagram}%
Schematic illustration of the experimental setup. CDS: cylindrical detector system, CDH: cylindrical scintillator hodoscope, CDC: cylindrical drift chamber, D$_2$ TGT: liquid deuterium target, T0: time zero counter, BPD: backward proton detector, BPC: backward proton drift chamber, DEF: beam defining counter, FDC: forward drift chamber, NC: neutron counter array, CVC: charged particle veto counter, and PC: proton counter array. The reaction diagram expected for $d(K^-,N)\pi\Sigma$ is shown in the inset.
}
\end{figure}

We carried out an experimental study of kaon-induced $\pi\Sigma$ production via $d(K^-,n)\pi\Sigma$ reactions \cite{e31}. Our expectation was to measure a reaction sequence consisting of a 1-GeV/$c$ incident negative kaon knocking out a neutron at a very forward angle (less than 6 degrees in the laboratory frame) from a deuteron, with the $\bar{K}$ recoiled backward reacting with the residual nucleon ($N_2$) to produce $\pi$ and $\Sigma$, as shown in the reaction diagram in the inset of Fig.~\ref{fig:diagram}. 
In the second step of the reaction sequence, $\bar{K}N_2\rightarrow\pi\Sigma$ scattering takes place even below the $\bar{K}N$ mass threshold. Since the typical momentum of a recoiled $\bar{K}$ is as low as $\sim$250 MeV/$c$ for a $\pi\Sigma$ invariant mass of around 1405 MeV/$c^2$, $S$-wave scattering is expected to be dominant. We measured the $\pi\Sigma$ invariant mass spectra, from which we deduced the $\bar{K}N$ scattering amplitude in the $I$ = 0 channel.

The experiment was performed at the K1.8BR beam line \cite{k18br} of the Japan Proton Accelerator Research Complex (J-PARC). Negatively charged kaons delivered from K1.8BR were incident on a liquid deuterium (D$_2$) target of 125 mm 
thickness. The momentum of the incident kaons was analyzed by the K1.8BR-D5 magnetic spectrometer. 
A schematic layout of the experimental setup \cite{hashimoto15,sada16} is illustrated in Fig.~\ref{fig:diagram}.
A time-zero counter (T0), placed 1100 mm upstream from the D$_2$ target, defined a time origin triggered by the incident kaon for time-of-flight measurements of scattered particles.
A drift chamber (BPC) and scintillator hodoscopes (BPD) were placed 143.2 mm and 482.5 mm upstream from the D$_2$ target center, respectively. Kaon beam tracks were measured by the BPC, which were used to determine the reaction vertex. The BPD and the BPC were used to detect backward emitted protons from $\pi^0\Sigma^0$ productions, as mentioned later. The kaon beam was finally defined by a beam defining counter (DEF) placed just in front of the D$_2$ target.
The integrated luminosity of the kaon beam used in the present analysis was ($5927\pm158$) [($8078\pm248$)] $\mu$b$^{-1}$, which was the product of the beam intensity 5.56$\times$10$^{10}$, the number of target deuterons 4.82$\times$10$^{23}$ [6.03$\times$10$^{23}$], efficiencies of the beam line detectors (28.1$\pm$1.7)\% [(31.5$\pm$1.9)\%)], trigger system (95.2$\pm$2.6)\%, and the data acquisition system (76.5$\pm$5.9)\%. Here, the numbers in square brackets are for the $\pi^0\Sigma^0$ mode.
Charged particles from the D$_2$ target were measured by a cylindrical detector system (CDS), consisting of a cylindrical drift chamber (CDC) and scintillator hodoscopes (CDHs) surrounding the D$_2$ target. An efficiency of the CDC for one-charged particle tracking was estimated to be (97.7$\pm$0.4)\%. 
The CDS was operated in a solenoid magnet with a magnetic field of 0.714 Tesla. 
Scattered neutrons were detected by neutron counters (NCs), consisting of an array of 112 plastic scintillator slabs (200 mm width, 1500 mm height, and 50 mm thickness each), placed approximately 15 m from the D$_2$ target. Since the solid angle of the NCs seen from the target is (21.5$\pm$0.2)\% msr, the angular coverage for the emitted neutrons is less than 6 degrees. 
The detection efficiency of the NC was estimated to be (31.7$\pm$1.6)\%. It was measured by finding a neutron at the NC to the predicted neutron emission direction in the $p(K^-,\bar{K}^0)n$ reaction, where a backward-recoiled $\bar{K}^0$ was reconstructed by the CDS. In reality, a factor of (91.9$\pm$0.7)\% due to a charged-particle veto counter (CVC) placed in front of the NC to veto charged particles, was multiplied as an effective efficiency of the NC. 
Charged particles emitted in a forward angle, including the incident beams, were swept out by a dipole magnet placed behind the solenoid magnet. Protons knocked out from deuterons by the incident kaon beam were bent by the dipole magnet in the opposite direction of the beam. The time-of-flight of the knocked-out proton was measured by proton counters (PC), consisting of hodoscopes of 27 scintillator slabs (100 mm width, 1500 mm height, and 50 mm thickness each), which were placed beside the CVC. The solid angle of the PC is slightly momenum dependent and is typically 22.6 msr. The trajectory of each scattered proton was determined using the position information from the reaction vertex at the target, a drift chamber (FDC) placed at the entrance of the dipole magnet, and the hit slab of the PC. 
A tracking efficiency for the proton is estimated to be (81.9$\pm$4.2)\%. 
We measured the $d(K^-,p)\pi^-\Sigma^0$ reaction, where the $p$ was detected by the PC and the two $\pi^-$s were detected by the CDS.

\begin{figure}
\centering
\includegraphics[width=7.8cm]{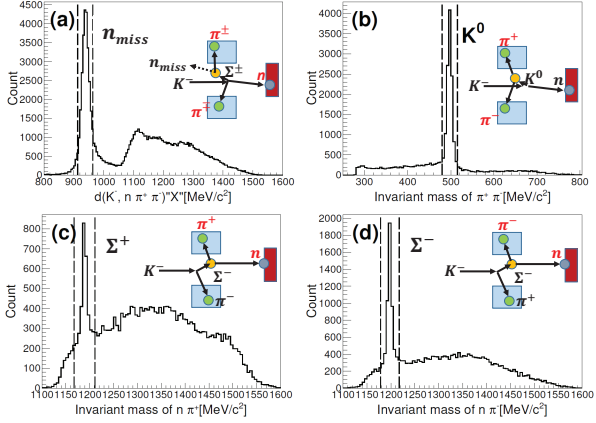}%
\caption{\label{fig:missing-n}%
(a) Missing mass spectrum of $d(K^-,n\pi^+\pi^-)$ togather with an illustration of a typical signal event topology. The missing neutron ($n_{\rm miss}$) is identified separately. The dashed lines indicate a selected neutron mass region. (b), (c), and (d) Invariant mass spectra of $\pi^+\pi^-$, $n\pi^+$, and $n\pi^-$, respectively, in the $d(K^-,n\pi^+\pi^-)n_{\rm miss}$ reactions. The peak regions shown with the dashed lines were excluded as $K^0$  and one-step $\pi^\mp\Sigma^\pm$ production processes as typical event topologies are illustrated, respectively.}
\end{figure}

\begin{figure}[t]
\centering
\includegraphics[width=7.8cm]{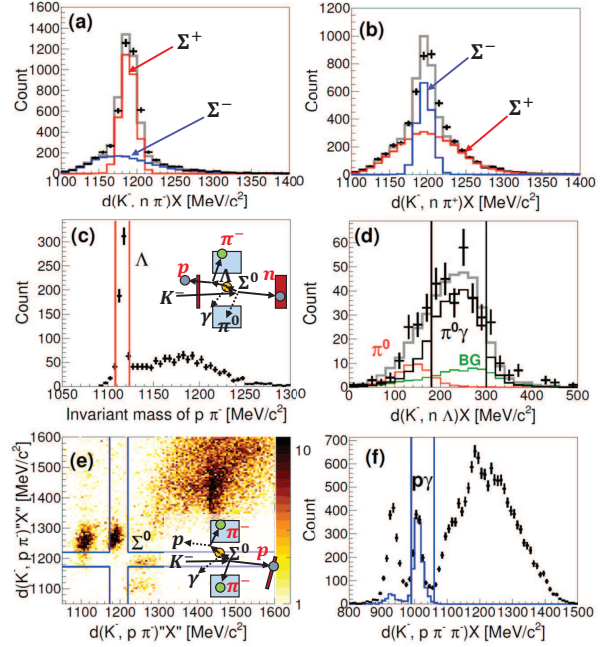}%
\caption{\label{fig:mode-id}%
(a) and (b) Decomposed $\Sigma^\pm$ peaks in the $d(K^-,n\pi^\mp)$ missing mass spectra, respectively. (c) Invariant mass spectrum of $\pi^-$ and $p$ measured by CDS and BPD/BPC, respectively. A $\Lambda$ peak was selected for the $\pi^0\Sigma^0$ mode (vertical lines) as the $\Sigma^0$ immediately decays into $\Lambda+\gamma$. A typical event topology for the $\pi^0\Sigma^0$ mode is illustrated in the figure.
(d) Missing mass spectrum of $d(K^-,n\Lambda)$. 
The expected $\pi^0$, $\pi^0\gamma$, and background components (BG) are overlaid as histograms. See text for BG. A $\pi^0\gamma$ region (0.18--0.3 GeV/$c^2$) was gated for the $\pi^0\Sigma^0$ mode. 
(e) Scatter plot of two possible $d(K^-,p\pi^-)$ missing masses for the $\pi^-\Sigma^0$ mode. A typical event topology for the $\pi^-\Sigma^0$ mode is illustrated in the figure. A $\Sigma^0$ mass region was selected, as indicated by the blue lines. (f) Missing mass spectrum of $d(K^-,p\pi^-\pi^-)$ (crosses) and selected $\Sigma^0$ region in (e) (histogram). A $p\gamma$ peak was selected to identify the $\pi^-\Sigma^0$ mode (vertical lines). 
}
\end{figure}

We measured the $\pi^\pm\Sigma^\mp$ production associated with a knocked-out neutron detected by the NC, where $\pi^+$ and $\pi^-$ were detected by the CDS and the missing neutron was identified separately in a $d(K^-,n\pi^+\pi^-)$ missing mass spectrum, as shown in Fig.~\ref{fig:missing-n}. In these modes, three background processes are relevant as they all give the same final state of $n\pi^+\pi^-n_{\rm miss}$, where $n_{\rm miss}$ represents the neutron identified in the $d(K^-,n\pi^+\pi^-)$ missing mass spectrum: (1) $K^-d\rightarrow n\bar{K}^0n_{\rm miss}$, (2) $K^-d\rightarrow\pi^-\Sigma^+n_{\rm miss}$, and (3) $K^-d\rightarrow\pi^+\Sigma^-n_{\rm miss}$. In (2) and (3), $\pi^\mp\Sigma^\pm$ are produced with an incident $K^-$ interacting with a bound proton in a deuteron. A neutron from the $\Sigma$ decay is emitted at a forward angle and detected by the NC. The $n_{\rm miss}$ is a spectator neutron in the processes. They are the so-called one-step $\pi^\mp\Sigma^\pm$ production processes, which occurs in a quite differrent kinematical region compared with that for the two-step process that we concern in the present article. Above three processes can be excluded since we can identify the $\bar{K}^0$, $\Sigma^+$, and $\Sigma^-$ peaks in the invariant mass spectra of $\pi^+\pi^-$, $n\pi^+$, and $n\pi^-$, as shown in Fig.~\ref{fig:missing-n}(b), (c), and (d), respectively.
We obtained the $\pi^\pm\Sigma^\mp$ missing mass spectra in the $d(K^-,n){\pi^\pm\Sigma^\mp}$ reactions separately, as we will show later. 
The production ratio of $\pi^-\Sigma^+$ to $\pi^+\Sigma^-$ was obtained to reproduce the $\Sigma^\pm$ peak and its kinematic reflection (continuum-like distribution). The decomposed $d(K^-,n\pi^\pm)$ missing mass spectra are shown in Fig.~\ref{fig:mode-id}(a) and (b), respectively.

In the $\pi^0\Sigma^0$ production, $\Sigma^0$ immediately decays to $\Lambda\gamma$. The $\Lambda$ hyperon decays to $\pi^-$ and a proton. The $\pi^-$ is emitted in a wide angular region and could be detected by the CDS. While, the proton is generally emitted backward because most of the momentum of the $\pi^0\Sigma^0$ that recoiled backward in the $d(K^-,n)$ reaction is carried by the heavier particle. We measured the time-of-flight of the backward proton, detected by the BPC and the BPD. We identified the decaying $\Lambda$ in the invariant mass spectrum reconstructed from the measured momenta of the $\pi^-$ and the proton [Fig.~\ref{fig:mode-id}(c)]. Then, the missing mass spectrum of $d(K^-,n\Lambda)$ was obtained as shown in Fig.~\ref{fig:mode-id}(d). 
The missing $\pi^0$, $\pi^0\gamma$, and background components (BG) 
contributions were decomposed based on a Monte Carlo simulation, as indicated in the figure, which were 12\%, 70\%, and 18\%, respectively. 
Here, hyperon ($Y$)-production processes that associate with a backward proton ($K^-d\rightarrow p(Y\pi)^-$) and those induced by quasi-free backward kaons that react with an another deuteron ($d^\prime$) in the deuterium target ($K^-d\rightarrow\bar{K}X, \bar{K}d^\prime\rightarrow YX^\prime$) are taken into accout as the BG components.
By gating the mass window for 0.18 to 0.3 GeV/$c^2$ in the spectrum, we obtained the $\pi^0\Sigma^0$ mode with only a small amount of contamination from the $\pi^0\Lambda$ mode and background components, which were reduced to be 1.0\% and 3.9\%, respectively. 
The contribution of the contamination is subtracted in the present $\pi^0\Sigma^0$ missing mass spectrum.

The $\pi^-\Sigma^0$ mode was identified by selecting the $\Sigma^0$ and $p\gamma$ mass regions in a scatter plot of the two possible $d(K^-,p\pi^-)$ missing masses and the $d(K^-,p\pi^-\pi^-)$ missing mass spectrum, as shown in Fig.~\ref{fig:mode-id}(e) and (f), respectively. The missing $p\gamma$ mass distribution is isolated since $\Sigma^0$ is moving slowly.

\section{$\pi\Sigma$ mass spectra}

\begin{figure}
\centering
\includegraphics[width=7.7cm]{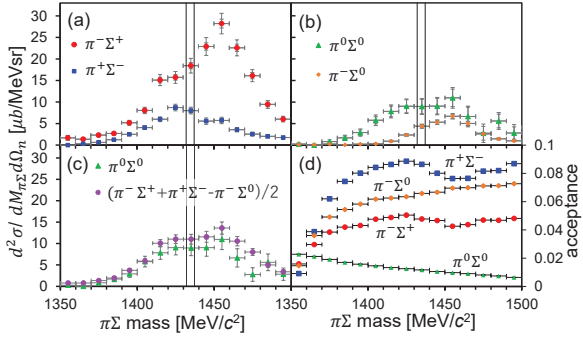}%
\caption{\label{fig:spectra}%
Measured spectra of (a) $\pi^\pm\Sigma^\mp$, (b) $\pi^0\Sigma^0$ and $\pi^-\Sigma^0$, and (c) $\pi^0\Sigma^0$ and $(\pi^+\Sigma^-+\pi^-\Sigma^+-\pi^-\Sigma^0)/2$. (d) Acceptances of the detectors for $\pi^\pm$, $\pi^-p$, and $2\pi^-$ from $\pi^\pm\Sigma^\mp$, $\pi^0\Sigma^0$, and $\pi^-\Sigma^0$, respectively. 
Statistical and total errors are shown separately as inner and outer bars in (a) and (b), while only total errors are shown in (c). 
The vertical thin lines shows the $K^-p$ and $K^0n$ mass thresholds.}
\end{figure}

The mass spectra of $\pi^\pm\Sigma^\mp$, $\pi^0\Sigma^0$, and $\pi^-\Sigma^0$ were obtained, as shown in Fig.~\ref{fig:spectra}. 
Errors in the vertical axes include statistical errors, scaling factor errors, and systematic uncertainties. The scaling factor errors arise from uncertainties in corrections of the target thickness and beam intensity, the efficiencies of the data acquisition system, detectors, and event selections in the analysis codes, and the geometrical acceptances and efficienceies of the relevant particle detectors. Geometrical acceptances of the detector setup for $\pi^\pm$, $\pi^-p$, and $2\pi^-$ from $\pi^\pm\Sigma^\mp$, $\pi^0\Sigma^0$, and $\pi^-\Sigma^0$, respectively, are shown in Fig.~\ref{fig:spectra}(d), which were evaluated by Monte Carlo simulations in conditions that knocked-out neutron and proton are detected at the NC and the PC, respectively. The scaling factor errors relative to the obtained cross sections for $\pi^{\pm}\Sigma^{\mp}$, $\pi^0\Sigma^0$, and $\pi^-\Sigma^0$ are estimated to be 5.8\%, 6.2\%, and 3.5\%, respectively. 
For the $\pi^\pm\Sigma^\mp$ spectra, the fitting errors to separate the two modes as described in Fig.~\ref{fig:mode-id}(a) and (b) are also taken into account, which are dominant sources of systematic uncertainties in estimations of the cross sections. The fitting errors relative to the cross sections are typically 7\% and 6\% at the $K^-p$/$K^0n$ mass thresholds for the $\pi^\pm\Sigma^\mp$ modes, respectively. 
In the case of the $\pi^0\Sigma^0$ mode, the fitting error to decompose the $\pi^0\Lambda$ mode and the other background mentioned in the previous section is dependent on the missing mass. It is typically 1.5\% at the $\bar{K}N$ mass threshold. On the other hand it is 13\% at around 1475 MeV/$c^2$, where contamination of the BG conponents is maximum.
The statistical and total errors are shown separately as inner and outer bars in Fig.~\ref{fig:spectra}(a) and (b), while only the total errors are shown in Fig.~\ref{fig:spectra}(c).

We observed different line shapes in the $\pi^\pm\Sigma^\mp$ modes [Fig.~\ref{fig:spectra}(a)]. Since both the $I$ = 0 and 1 amplitudes contribute to the modes, the difference is due to interference between the two amplitudes. In the $\pi^-\Sigma^+$ mode, we find a bump around 1450 MeV/$c^2$ with a small shoulder below the $K^-p$ mass threshold. On the other hand, the $\pi^+\Sigma^-$ spectrum shows a broad distribution with a maximum strength just below the $K^-p$ mass threshold. 
The $\pi^0\Sigma^0$ and $\pi^-\Sigma^0$ modes [Fig.~\ref{fig:spectra}(b)] contain only the $I$ = 0 and 1 amplitudes, respectively. 
The strength of the $\pi^-\Sigma^0$ spectrum is smaller than that of the $\pi^0\Sigma^0$ spectrum. We find that the $I$ = 0 amplitude is dominant, particularly below the $\bar{K}N$ mass threshold. We find no structure at around 1385 MeV/$c^2$ in the $\pi^-\Sigma^0$, where we might expect a structure of the $\Sigma^*(1385)$ resonance. This fact suggests dominance of $S$-wave $\pi\Sigma$ production in the present reactions, since $\Sigma^*(1385)$ decays into a $P$-wave $\pi\Sigma$ state.

The $\pi\Sigma$ production cross sections can be described with $T_1^I$ and $T_2^{I^\prime}$ as follows;
\begin{eqnarray}
\frac{d\sigma}{d\Omega}(\pi^\pm\Sigma^\mp)\propto
\left|C_1^0T_2^{I^\prime=0}\mp{C_1^1T_2^{I^\prime=1}}\right|^2,\\
\frac{d\sigma}{d\Omega}(\pi^-\Sigma^0)\propto\left|C_1^1T_2^{I^\prime=1}\right|^2,\\
\frac{d\sigma}{d\Omega}(\pi^0\Sigma^0)\propto\left|C_1^0T_2^{I^\prime=0}\right|^2,\\
C_1^0=\frac{3T_1^{I=0}-T_1^{I=1}}{4\sqrt{3}},\ \ 
C_1^1=\frac{T_1^{I=1}+T_1^{I=1}}{4}.
\end{eqnarray}
Here, $T_1^I$ and $T_2^{I^\prime}$ represent the scattering amplitude of the first-step and second-step two-body $K^-N_1\rightarrow\bar{K}N$ and $\bar{K}N_2\rightarrow\pi\Sigma$ reactions with isospin $I$ and $I^\prime$, respectively. 
The coefficients are determined by the sums of the products of the Clebsch--Gordan coefficients in terms of the isospin in the possible processes in the two-step reaction, as described as follows:
\begin{eqnarray}
\sum_{m_X,I,m,I^\prime,m^\prime}{
\langle\frac{1}{2}m_{N_1}\frac{1}{2}m_{N_2}|00\rangle\langle\frac{1}{2}m_{\bar{K}}%
\frac{1}{2}m_{N}|Im\rangle}\nonumber\\
\times\langle\frac{1}{2}m_{K^-}\frac{1}{2}m_{N_1}|Im\rangle{T_1^I} \nonumber\\
\times\langle\frac{1}{2}m_{\pi}\frac{1}{2}m_{\Sigma}|I^\prime{m^\prime}\rangle%
\langle\frac{1}{2}m_{\bar{K}}\frac{1}{2}m_{N_2}|I^\prime{m^\prime}\rangle{T_2^{I^\prime}},
\end{eqnarray}
where $m_{X=N_1, N_2, \bar{K}, N, K^-, \pi, \Sigma}$ is a $z$-component of the isospin of a relevant particle $X$. 
Then, one finds a relation among the four reaction cross sections as
\begin{eqnarray}
\frac{1}{2}\frac{d\sigma}{d\Omega}(\pi^+\Sigma^-+\pi^-\Sigma^+-\pi^-\Sigma^0)=\frac{d\sigma}{d\Omega}(\pi^0\Sigma^0).
\end{eqnarray} 
We confirmed the relationship, as demonstrated in Fig.~\ref{fig:spectra}(c). 

\section{Discussion}

\begin{figure}
\centering
\includegraphics[width=7cm]{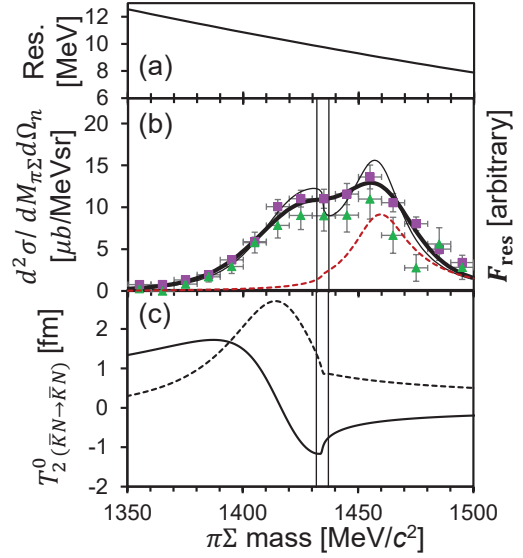}%
\caption{\label{fig:fitting}%
(a) Experimental resolution as a function of the $\pi\Sigma$ mass. 
(b) Calculated $\pi\Sigma$ spectrum to fit the measured spectra in the $I$ = 0 channel. The solid thick and thin lines are the spectrum with and without the resolution function convoluted, respectively. The response function $F_{\rm res}$ is shown as a dashed line in arbitrary units. (c) Deduced scattering amplitude of $\bar{K}N\rightarrow\bar{K}N$ in the $I$ = 0 channel. The real and imaginary parts are shown as solid and dashed lines, respectively. The vertical thin lines show the $K^-p$ and $K^0n$ mass thresholds.
}
\end{figure}

Several authors have discussed $\pi\Sigma$ production associated with nucleon emission in kaon induced reactions on deuterons \cite{jido2,yamagata,miyagawa,miyagawa2,kamano}, and hence we describe the $\pi\Sigma$ spectral shape assuming that the two-step reaction 
is dominant when the knocked-out nucleon is emitted at a very forward angle. We neglect the direct production of $\pi\Sigma$ by collisions of incident $K^-$ with nucleons in deuteron as its contribution is negligibly small at the very forward angle of knocked-out neutron.
Then, the $\pi\Sigma$ production cross section can be described as
\begin{eqnarray}
\frac{d^2\sigma}{dM_{\pi\Sigma}d\Omega_n}\sim\left|\langle{n\pi\Sigma}|T_2G_0(\bar{K},N_2)T_1|K^-\Phi_d\rangle\right|^2,\\
T_2=T_2^{I^\prime}(\bar{K}N_2,\pi\Sigma),\\
T_1=T_1^I(K^-N_1,\bar{K}N),
\end{eqnarray}
where $|K^-\Phi_d\rangle$ and $|n\pi\Sigma\rangle$ denote the initial $K^-$ and deuteron and final $n\pi\Sigma$ wave functions, respectively. 
$G_0(\bar{K},N_2)$ is the Green's function which describes the intermediate $\bar{K}$ propagation between the two vertices. More detailed expressions can be found in Refs.~\cite{miyagawa,jido2,kamano}.
The cross section can be simplified by a factorization approximation, as follows:
\begin{eqnarray}
\frac{d^2\sigma}{dM_{\pi\Sigma}d\Omega_n}\approx\left|T_2^{I^\prime}\right|^2F_{\rm res}(M_{\pi\Sigma}),\\
F_{\rm res}(M_{\pi\Sigma})=\left|\int{G_0T_1^I\Phi_d(q_{N_2})d^3q_{N_2}}\right|^2.
\end{eqnarray}
Here, $q_{N_2}$ is the momentum of the residual nucleon.
In this way, the $\pi\Sigma$ spectrum can be decomposed into $T_2^{I^\prime}$ and the response function $F_{\rm res}$. Using the $K^-N\rightarrow\bar{K}N$ scattering amplitudes based on a partial wave analysis \cite{gopal} and the deuteron wave function $\Phi_d$ \cite{deuteron}, we evaluate $F_{\rm res}$ as a function of the $\pi\Sigma$ mass $M_{\pi\Sigma}$, as shown by the dashed line in Fig.~\ref{fig:fitting}(b). Here, we took 3 degrees as a typical scattering angle of the knocked-out 
nucleon in the laboratory frame. 
The line shapes of the $\pi\Sigma$ mass spectra above the $\bar{K}N$ mass threshold are characterized by $F_{\rm res}$, the distribution of which reflects the Fermi motion of a nucleon in the dueteron. 
For $S$-wave $T_2^{I^\prime}$, we consider the $\bar{K}N$-$\pi\Sigma$ coupled channel $T$ matrix. The diagonal and off-diagonal matrix elements can be parametrized similarly to the case in Ref.~\cite{lensniak} as
\begin{eqnarray}
T_2^{I^\prime}(\bar{K}N,\bar{K}N)=\frac{A^{I^\prime}}{1-iA^{I^\prime}k_2+\frac{1}{2}A^{I^\prime}R^{I^\prime}k_2^2},\\
T_2^{I^\prime}(\bar{K}N,\pi\Sigma)=\frac{e^{i\delta^{I^\prime}}}{\sqrt{k_1}}\frac{\sqrt{{\rm Im}A^{I^\prime}-\frac{1}{2}|A^{I^\prime}|^2{\rm Im}R^{I^\prime}k_2^2}}{1-iA^{I^\prime}k_2+\frac{1}{2}A^{I^\prime}R^{I^\prime}k_2^2},
\end{eqnarray}
where $A^{I^\prime}$, $R^{I^\prime}$, and $\delta^{I^\prime}$ are the complex scattering length, complex effective range, and real phase, respectively. 
$k_1$ and $k_2$ are respectively the momenta of $\pi$ and $\bar{K}$ in the center of mass frame. Here, $k_2$ becomes a pure imaginary number below the $\bar{K}N$ mass threshold, to satisfy analytic continuity.

We demonstrate the fitting result for the $\pi\Sigma$ ($I$ = 0) channel, as shown in Fig.~\ref{fig:fitting}(b). $A^0$ and $R^0$ are determined to fit the measured $\pi^0\Sigma^0$ and $(\pi^+\Sigma^-+\pi^-\Sigma^+-\pi^-\Sigma^0)/2$ spectra, simultaneously. We took the $\bar{K}N$ mass threshold at the average of $K^-p$ and $K^0n$ since the differential cross sections of $K^-n\rightarrow K^-n$ \cite{npb129} and $K^-p\rightarrow K^0n$ \cite{npb90} are almost equal at a neutron forward angle at an incident kaon momentum of $\sim$1 GeV/$c$. However, we took into account the differences from the fitting results for the cases of the $K^-p$ and $K^0n$ mass thresholds as systematic errors. In the present fitting, $\delta^{I^\prime}$ could not be determined since it deos not appear explicitly in the fitting function that depends on $|T_2^{I^\prime}(\bar{K}N,\pi\Sigma)|^2$. In the fitting, the experimental resolution function [Fig.~\ref{fig:fitting}(a)] was convoluted with the calculated spectrum and the vertical scale is arbitrarily adjusted. We obtained
$A^0 = [-1.12\pm 0.11(\mbox{fit})^{+0.10}_{-0.07}(\mbox{syst.})]$ + $[0.84\pm 0.12(\mbox{fit})^{+0.08}_{-0.07}(\mbox{syst.})]i$ fm, 
$R^0 = [-0.18\pm0.31\mbox{(fit)}^{+0.08}_{-0.06}(\mbox{syst.})]$ + $[-0.40\pm0.13\mbox{(fit)}\pm0.09(\mbox{syst.})]i$ fm,
where the fitting errors are indicated as ``(fit)''. As mentioned above, the differences of the different $\bar{K}N$ mass threshold were taken into account as systematic errors indicated as ``(syst.)''.  The reduced chi-square was 1.76 with 24 degrees of freedom. 
The present scattering length is smaller than a recent theoretical calculation, $-1.77+1.08i$, which is based on the lattice QCD \cite{liu}. 
The thick and thin solid lines in Fig.~\ref{fig:fitting}(b) show the resolution-convoluted and no-resolution-convoluted spectra, respectively, calculated with the best fit values. 
The energy dependence of the deduced 
$T_2^0(\bar{K}N,\bar{K}N)$ is shown in Fig.~\ref{fig:fitting}(c). 
We find a zero-crossing in the real part and a bump in the imaginary part at the same place. This is a typical structure of a resonance. We find a resonance pole at
$1417.7^{+6.0}_{-7.4}(\mbox{fit})^{+1.1}_{-1.0}(\mbox{syst.})$ + 
$[-26.1^{+6.0}_{-7.9}(\mbox{fit})^{+1.7}_{-2.0}(\mbox{syst.})]i$ MeV/$c^2$
in the $I$ = 0 channel of the $\bar{K}N\rightarrow\bar{K}N$ scattering. The errors are estimated by fluctuations of the pole position due to the errors for the best fit values of $A^0$ and $R^0$. The real part of the deduced pole is closer to the $K^-p$ mass threshold than the so-called PDG value of 1405.1 MeV/$c^2$. 
It is worthy of evaluating the following quantity, $|T_2^0(\bar{K}N,\bar{K}N)|^2/|T_2^0(\bar{K}N,\pi\Sigma)|^2\sim2.2^{+1.0}_{-0.6}$(fit)$\pm0.3$(syst.) at the pole energy, which corresponds to the ratio of the two partial widths in the Flatt{\'e} formula \cite{flatte,chung95}. This suggests that the coupling of $\Lambda(1405)$ to $\bar{K}N$ is predominant, which does not contradict a picture of $\Lambda(1405)$ as a $\bar{K}N$-bound state. 
Mei{\ss}ner and Hyodo have reviewed and discussed the pole structure of the $\Lambda(1405)$ region based on chiral unitary approaches with a constraint on the scattering length obtained from kaonic hydrogen atom $X$-ray data by the SIDDHARTA collaboration \cite{siddharta11,siddharta12}\cite{mh-pdg}. They collected four sets of two poles deduced by several authors in the relevant region. Poles 1 and 2 are the so-called higher and lower poles, respectively, which are thought to be coupled to $\bar{K}N$ and $\pi\Sigma$, respectively.
The suggested higher poles are located at the region of 1421--1434 MeV on the real axis and 10--26 MeV on the imaginary axis in the complex energy plane. The pole position determined by the present experiment is consistent to the higher poles though it is located at slightly smaller and larger values for the real and imaginary parts, respectively. A lattice QCD calculation has reported two poles and the so-called higher pole is located at $1430-22i$ MeV/$c^2$ \cite{liu2}. Our result is smaller and similar in real and imaginary part, respectively. Recently, Anisovich $et\ al.$ reported one single pole of $\Lambda(1405)$ contribution to fit the data of $\gamma$ and $K^-$ induced reactions on proton and the kaonic hydrogen atom, as $1422\pm 3-(21\pm 3)i$ MeV/$c^2$ \cite{anisovich}. The present result is consistent with the reported pole position.

\section{Conclusion}

We measured $\pi^\pm\Sigma^\mp$, $\pi^0\Sigma^0$, and $\pi^-\Sigma^0$ mass spectra below and above the $\bar{K}N$ mass threshold in $d(K^-,N)\pi\Sigma$ reactions at a forward angle, of $N$ knocked out by an incident kaon momentum of 1 GeV/$c$. We obtained decomposed $\pi\Sigma$ spectra in terms of $I$ = 0 and 1, and confirmed a relation between the four reactions with respect to the isospin states. We find that the $I$ = 0 amplitude is dominant. We demonstrated that the $\pi\Sigma$ spectral shape in the $I$ = 0 channel is well reproduced by the two-step reaction of a neutron knocked out at a forward angle by an incident negative kaon and a recoiled $\bar{K}$ reacting with a residual nucleon in deuteron to produce $\pi\Sigma$ in the $I$ = 0 state. We deduced the two-body $\bar{K}N$ scattering amplitude in the $I$ = 0 channel around the $\bar{K}N$ mass threshold, from which we find a resonance pole at 
$1417.7^{+6.0}_{-7.4}(\mbox{fit})^{+1.1}_{-1.0}(\mbox{syst.})$ + 
$[-26.1^{+6.0}_{-7.9}(\mbox{fit})^{+1.7}_{-2.0}(\mbox{syst.})]i$ MeV/$c^2$. 
The present data provide fundamental information on the $\bar{K}N$ interaction and kaonic nuclei \cite{kpp,kpp2}. 

\section*{Acknowledgements}
The authors would like to express their thanks to the J-PARC PAC members and the crews of the J-PARC accelerator and hadron facility group for their encouragement, support, and stable delivery of beams for the E31 experiment. We are grateful to Professor D. Jido, Dr. T. Sekihara, and Professor J. Yamagata-Sekihara for their support since the planning stage of the E31 experiment. We are grateful to Professor K. Miyagawa and Dr. H. Kamano for their contributions to the calculations of the $\pi\Sigma$ spectral shapes. 
The present work was supported by MEXT Grants-in-Aid of Innovative Area No.\ 21105003, No.\ 18H05402, and a Grant-in-Aid of Scientific Research A No.\ 16H02188 and S No.\ 22H04940. 
%




\end{document}